# Addressing band-edge-property spatial variations and localized-state carrier trapping and recombination in solar cell numerical modeling

Yiming Liu[1], Stephen Fonash[2]

[1] Nankai University, 94 Weijin Rd., Tianjin, 300071, P.R. China

[2] Center for Nanotechnology Education and Utilization, The Pennsylvania State University, University Park, PA, 16802

**ABSTRACT**

Conduction and valence band-edge-property variations with position as well as defects giving rise to localized states in the energy gap can play a significant role in determining solar cell performance. Understanding their effects on a device is necessary in interpreting complex experimental observations and in optimizing the performance of solar cells. In this overview, we include the effective forces arising from electron and hole band-edge-property variations with position in a numerical formulation of solar cell performance. Further we systematically catalogue and review a variety of localized states with different types and distributions, and include in our numerical transport model the carrier trapping, electric field modification, and recombination caused by these localized states. The successful implementation of the numerical modeling of band-edge-property variations and defect state effects is demonstrated using the methodology of the solar cell simulation code Analysis of Microelectronic and Photonic Structures (AMPS) and its derivatives.

1. INTRODUCTION

Among the many factors that influence solar cells, band-edge-property (energy position and effective density of states) variation with position and the properties of localized states (density, energy position, capture cross-sections, etc.) can play an important part in determining device characteristics. Variations in the electron and hole band-edge-properties with position can arise inadvertently due to atomic interdiffusion at interfaces or purposefully at contacts and heterojunctions. Generally analytical and most numerical codes used for solar cell modeling neglect the effective forces arising from band-edge-property variations, yet these can have magnitudes of the order of those of electric field forces.

Dopants are purposefully present gap states; defect states are not. Defect states, which may arise from a variety of sources including structure and impurities, may be classified into three types: Urbach band tail states, discrete gap energy states, and banded gap localized states. Gap defect states may be donor- or acceptor-like single particle states or they may be multi-particle amphoteric states [1]. Gap states may be functions of position in bulk materials and be very different in their properties at interfaces. The importance of localized states to device performance lies in their being able to trap holes and electrons thereby giving rise to recombination, trapping, and space charge. These possibilities can not only affect the cell bulk region but can have a significant impact on interface recombination and fields at device contacts and heterojunctions. Analytical modeling and some numerical codes treat bulk recombination with linearized carrier lifetime models and treat contact and interface recombination with linearized recombination rate models. Most neglect the implications of localized space charge on contact barriers, bulk electric field and transport, assume full ionization of dopants, and neglect the possibility of amphoteric behavior [1].

All of these approximations become increasingly problematic as the community moves to further developing advanced solar call structures. However, with today's computing power, these approximations can be easily avoided with numerical modeling that is rapid and user-friendly. The full treatment of the impact of both band-edge-property variations and localized states was first done in the Analysis of Microelectronic and Photonic Structures (AMPS) computer code [2-4] and this comprehensive treatment is also found in the several AMPS derivatives now in use [5, 6]. With the increasing utilization of computational tools for the development of solar cells heavily influenced by gap states, there is a need for the review presented here to enable one to compare the comprehensiveness of available numerical programs. To further assist this comparison, the successful implementation of the numerical modeling of effective forces and defect state effects is discussed using the methodology of the AMPS family of computer modeling tools.

2. **TRANSPORT MODEL**

Transport in AMPS [2] and in its derivative codes [5, 6] is described in the domain defining the interior of a device by the continuity equations for electrons and holes as well as by Poisson's equation. The current densities required in the continuity equations are modeled using the drift-diffusion picture, generalized to include effective forces arising from spatial variations in conduction band edge effective density of states and energy $E_c$ and valence band edge effective density of states and energy $E_v$ [1]. Use of the drift-diffusion transport picture is valid so long as carriers' mean free path lengths are less than the characteristic device dimensions [1]. Poisson's equation also requires the band (including Urbach tail) and gap state populations. These are obtained by using the Shockley-Read-Hall (SRH) interband traffic formulation [1]. At the boundaries of this domain, which are the contacts since AMPS and its derivatives are 1-D programs, boundary conditions are imposed to complete the mathematical system. These boundary conditions are on the local vacuum level $\psi$ and on the electron $J_n$ and hole $J_p$ current densities.

The conditions imposed on the local vacuum level $\psi$ are (1) that $\psi = 0$ at one boundary and (2) that $\psi = (\psi_0 - V)$ at the other boundary where $\psi_0$ is its value at thermodynamic equilibrium and $V$ is the voltage. These conditions are of very general validity [1]. The current density boundary conditions are defined in terms of the standard surface recombination speed model which gives current densities that depend linearly on free carrier populations [1, 7]. Importantly, the use of this linear modeling at contacts for current densities does not limit the generality of the AMPS family of codes since the boundary conditions are only used to terminate the mathematical domain. AMPS actually allows for very non-linear phenomena at contacts since a layer with band-edge-properties different from the adjacent bulk material or a defect-laden layer, with any of the rich variety of recombination processes discussed below, or a layer with both attributes may be imposed contiguous to a contact. Such a layer can be in control of the contact recombination, barrier shape and charge trapping.

The mathematical system of Poisson's equation, continuity equations, generalized drift-diffusion current density models, free carrier population statistics, localized state population statistics, and the boundary conditions can be consolidated down to three equations plus boundary conditions by substituting the population and current density models into Poisson's and the two continuity equations. If the system is constrained to steady-state situations, the resulting three equations are the following:

$$\frac{d}{dx}(\varepsilon \frac{d}{dx}\psi) - q(p - n + N_d^+ - N_a^- + p_t - n_t) = 0 \qquad (1)$$

$$\frac{dJ_n}{dx} - q(R - G) = 0 \qquad (2)$$

$$\frac{dJ_p}{dx} - q(G - R) = 0 \qquad (3)$$

Here $\varepsilon$ is the permittivity, which may be a function of position, $q$ is the magnitude of the charge on an electron, and $G$ and $R$ are free carrier photogeneration and recombination at per volume per time, respectively. The current densities and concentrations of free holes $p$, free electrons $n$, ionized donor dopants $N_d^+$, ionized acceptor dopants $N_a^-$, holes trapped in defects $p_t$, and electrons trapped in defects $n_t$ which appear in the equations are all dependent on $\psi$, $E_{fn}$ and $E_{fp}$ at the position coordinate $x$. The rates of net defect-assisted and band-to-band recombination $R$ are dependent on $\psi$, $E_{fn}$ and $E_{fp}$ at the position coordinate $x$ whereas the electron current density $J_n$, and the hole current density $J_p$ are all also dependent on $\psi$, $E_{fn}$ and $E_{fp}$ and their derivatives at the position coordinate $x$. The optical generation $G$ only depends on $x$. In AMPS, it is obtained from the Beer-Lambert law with allowance for internal interface reflection but not interference effects [1]. In wxAMPS, $G$ can be optionally uploaded from other optical models and therefore can be generated by physical optics codes.

## 3. EFFECTIVE FORCES

As noted, variations in conduction band edge energy $E_c$ and the conduction band effective density of states $N_c$ give rise to effective forces acting on free electrons. These can be on a par with the force exerted on an electron by an electric field. In a comprehensive drift-diffusion model, the effective forces are incorporated in the electron current density expression via the second and third terms in the square bracket of Eq (4) below [1]; i.e.,

$$J_n = q\mu_n n [\xi - \frac{d\chi}{dx} - kT\frac{d\ln N_c}{dx}] + qD_n \frac{dn}{dx} \qquad (4)$$

Here $\xi$ is the electrostatic field, $\chi$ is the electron affinity, which locates $E_c$ with respect to the local vacuum level $\psi$, $\mu_n$ and $D_n$ are the electron mobility and diffusivity, respectfully, and $n$ is the free electron population.

Variations in valence band edge energy $E_c$ and the valence band effective density of states $N_c$ give rise to effective forces acting on free holes. These too can be on a par with the force exerted on a hole by an electric field. In a comprehensive drift-diffusion model, these hole effective forces are incorporated in the current density expression via the second and third terms in the square bracket of Eq (5) [1]; i.e.,

$$J_p = q\mu_p p [\xi - \frac{d(\chi + E_g)}{dx} + kT\frac{d\ln N_v}{dx}] - qD_p \frac{dp}{dx} \qquad (5)$$

Eq (4) and (5) give the current density models required for comprehensive numerical modeling. These are functions of $\psi$, $E_{fn}$ and $E_{fp}$ through the carrier populations, as we will see, and through the fact that $\xi = d\psi/dx$.

## 4. POPULATIONS AND RECOMBINATION TRAFFIC

In following our plan of using Eq. (1)-(3) together with the boundary conditions to determine $\psi$, $E_{fn}$ and $E_{fp}$ as functions of position, we require formulations of $n$, $p$, $N_d^+$, $N_a^-$, $p_t$, $n_t$, and $R$, in terms of $\psi$, $E_{fn}$ and $E_{fp}$.

### 4.1 Free Carriers

The free carrier populations $n$ and $p$ can be written at $x$ as

$$n = N_C e^{-(E_C - E_{Fn})/kT} \qquad (6)$$

and

$$p = N_V e^{-(E_{Fp} - E_V)/kT}. \qquad (7)$$

These expressions are valid in or out of thermodynamic equilibrium, and can be rewritten entirely in terms of our state variables $\psi$, $E_{fn}$, and $E_{fp}$ by choosing the Fermi level in the back contact at $x=L$ as the reference for these quantities. We take this contact to be a metal. Consequently its Fermi level is always separated in energy from the local vacuum level by the metal work function $\phi_o$ at x=L [1]. In that case the conduction band edge $E_c$ and the valence band edge $E_V$ at any point x in a device may be written as $E_c = \phi_o + \psi - \chi$ and $E_v = \phi_0 + \psi - \chi - E_g$, where $\chi$ is the affinity at x, and $E_g$ is the energy gap at point x. Using these expressions in the equations for n and p then gives statements written in terms of the variables $\psi$, $E_{fn}$ and $E_{fp}$ as well as written in terms of material parameters such as $N_c(x)$, $N_v(x)$, $\chi(x)$ and $E_g(x)$. While these expressions for n and p assume Boltzmann distributions, we will see that the Boltzmann formulation is not used to determine the populations $N_d^+, N_a^-$, $p_t$ and $n_t$ [1]. This allows the resulting expressions for these populations to be valid even when the defect state population and density of states may be comparable, as is very possible for gap states. One other comment is appropriate here: While AMPS takes the contacts at each boundary to be metals, there is no loss of generality since the layer adjacent to the contacts can be defined to be the actual contact; e.g., such layers could be defined to be transparent conducting oxides (TCO's) in the AMPS input.

**4.2 Gap States: General formulations**

Recombination in a device may be of three types: Shockley-Read-Hall, band-to-band, and Auger recombination. Only S-R-H recombination utilizes the gap states and therefore only it determines trapping, gap state charge, and SRH contributions to recombination. From the SRH model, the steady-state, localized-state-assisted recombination traffic R (carriers per time per volume) through $N_t$ states per volume at a discrete energy level E is [1, 8, 9]

$$R(E) = \frac{(np - n_i^2)V_{th}N_t\sigma_n\sigma_p}{\sigma_n(n + N_c e^{\frac{E-E_c}{\kappa T}}) + \sigma_p(p + N_v e^{\frac{E_v-E}{\kappa T}})} \quad (8)$$

where $V_{th}$ is the free carrier thermal velocity. The quantities $N_t$, $\sigma_n$, the capture cross-section of these states for free electrons, and $\sigma_p$, the capture cross-section of these states for free holes, may be functions of E and x. If the thermal velocity is not the same for free electrons and holes, the capture cross-sections may be appropriately adjusted to correct for this [1].

The derivation that leads to Eq. (8) may be used to show that the probability, $f_A(E)$, that these discrete states at energy E are occupied by an electron is given by [1, 8-10]

$$f_A(E) = \frac{\sigma_n n + \sigma_p N_v e^{\frac{E_v-E}{\kappa T}}}{\sigma_n(n + N_c e^{\frac{E-E_c}{\kappa T}}) + \sigma_p(p + N_v e^{\frac{E_v-E}{\kappa T}})} \quad (9)$$

whereas the probability $f_D(E)$ that they are occupied by a hole is $1 - f_A(E)$ or

$$f_D(E) = \frac{\sigma_p p + \sigma_n N_c e^{\frac{E-E_c}{\kappa T}}}{\sigma_n(n + N_c e^{\frac{E-E_c}{\kappa T}}) + \sigma_p(p + N_v e^{\frac{E_v-E}{\kappa T}})} \qquad (10)$$

These expressions are valid when the system is both out of thermodynamic equilibrium and at steady-state [1].

We note that band−to-band (or radioactive) recombination [1]

$$R(E) = \gamma(np - n_i^2) \qquad (11)$$

has been added along with S-R-H recombination in wxAMPS, in which the value of $\gamma$ is input directly. Band to band recombination does not utilize gap states and therefore, if present, only affects gap state populations indirectly through *n* and *p*. We also note that derivatives of the defect assisted recombination model and the band-to-band recombination model with respect to *ψ*, *Efn* and *Efp* are then used in Jacobian matrix element evaluations as discussed below.

### 4.3 Gap States: Dopant States

The number of ionized acceptor dopants $N_a^-$ at energy $E_a$ is the number of trapped electrons at these sites of density $N_a$ per volume. Consequently, $N_a^-$ is given by

$$N_a^- = f_A(E_a) N_a \qquad (12)$$

Since these states are acceptors, their contribution to the space charge is $-qN_a^-$ as seen in Eq. (1). It follows from our discussion of recombination that the steady-state traffic through these states is given by Eq. (8) with $N_t = N_a$.

The number of ionized donor dopants $N_d^+$ at energy $E_d$ is the number of trapped holes at these sites of density $N_d$ per volume. Consequently, $N_d^+$ is given by

$$N_d^+ = f_D(E_d) N_d \qquad (13)$$

Since these states are donors, their contribution to the space charge is $qN_d^+$ as seen in Eq. (1). It also follows from our discussion of recombination that the steady-state traffic through these states is given by Eq. (8) with $N_t = N_d$.

If there are donor states, acceptor states, or both present at various energies, multiple uses of these expressions must be made. Of course, one could always make the full ionization approximation in dealing with dopants in which case recombination traffic is neglected and the states are fully ionized [1].

However, in reality this may not be valid in a given materials system, and AMPS and its derivative codes give the user the ability to check on the appropriateness of the full ionization approximation. We note that it is the derivatives of these dopant state quantities with respect to ψ, $E_{fn}$ and $E_{fp}$ that are used in the Jacobian matrix element evaluations discussed below.

### 4.4 Gap States: Defect States:

#### 4.4.1 Discrete Defect States

In the simple situation of a defect having a discrete energy level $E_D$ in the band gap, the number of electrons per volume occupying these sites of density $N_{DD}$ is

$$n_{DD}(E_{DD}) = N_{DD} f_A(E_{DD}) \tag{14}$$

whereas the number of holes per volume occupying these sites is

$$p_{DD}(E_{DD}) = N_{DD} f_D(E_{DD}) \tag{15}$$

If these discrete defect states at energy $E_{DD}$ are acceptor-like, they contribute $n_{DD}(E_{DD})$ to the quantity $n_t$ in Eq. (1). If they are donor-like, they contribute $p_{DD}(E_{DD})$ to the quantity $p_t$ in Eq. (1). In either case, it follows from our discussion of recombination that the steady-state traffic through these states at energy $E_{DD}$ is given by Eq. (8) with $N_t = N_{DD}$; i.e., by

$$R_{DD}(E_{DD}) = \frac{(np - n_i^2) V_{th} N_{DD} \sigma_n \sigma_p}{\sigma_n (n + N_c e^{\frac{E_{DD} - E_c}{\kappa T}}) + \sigma_p (p + N_v e^{\frac{E_v - E_{DD}}{\kappa T}})} \tag{16}$$

Multiple uses of Eqs. (14)-(16) with different $N_{DD}$, $E_{DD}$, and capture cross-section values allow very general discrete defect state distributions to be constructed in the band gap. We note that it is the derivatives of these discrete defect state quantities with respect to ψ, $E_{fn}$ and $E_{fp}$ that are used in the Jacobian matrix element evaluations to be discussed below.

#### 4.4.2 Banded Defect States

Banded defect states are treated in AMPS and its derivatives as having a constant density of states per volume per energy $N_{BD}$ over an energy range of width $W$ centered at energy $E_{BD}$. The number of electrons occupying these states $n_{BD}(E_{BD})$ is given by

$$n_{BD}(E_{BD}) = N_{BD} \int_{E_{BD}-\frac{W}{2}}^{E_{BD}+\frac{W}{2}} f_A(E)dE \qquad (17)$$

and the number of holes occupying these states $p_{BD}(E_{BD})$ is given by

$$p_{BD}(E_{BD}) = N_{BD} \int_{E_{BD}-\frac{W}{2}}^{E_{BD}+\frac{W}{2}} f_D(E)dE \qquad (18)$$

If the states in a given band are acceptor-like, they contribute $n_{BD}(E_{BD})$ to the quantity $n_t$ in Eq. (1). Alternatively, if the states in a given band are donor-like, they contribute $p_{BD}(E_{BD})$ to the quantity $p_t$ in Eq. (1). In either case, it follows from Eq. (8) that the recombination traffic through a given band $R_{BD}(E_{BD})$ is given by

$$R_{BD}(E_{BD}) = N_{BD} \int_{E_{BD}-\frac{W}{2}}^{E_{BD}+\frac{W}{2}} \frac{(np-n_i^2)V_{th}\sigma_n\sigma_p dE}{\sigma_n(n+N_c e^{\frac{E-E_c}{\kappa T}}) + \sigma_p(p+N_v e^{\frac{E_v-E}{\kappa T}})} \qquad (19)$$

where AMPS and its derivatives take the cross-sections to be constant for a given band.

We note that it is the derivatives of these banded defect state quantities with respect to ψ, *E*fn and *E*fp that are used in the Jacobian matrix element evaluations discussed below. A demonstration of how terms are developed to allow these derivatives to be taken analytically is given in the **Appendix.** Multiple uses of Eqs. (17)-(19) with different $N_{BD}, E_{BD}$, and capture cross-section values allow very general bands of acceptor-like or donor-like defect states to be constructed in the band gap.

### 4.4.3    Gaussian Defect States

In the case of some materials, the energy distribution of at least some defects in the energy gap is more aptly described by a Gaussian density of states function. This continuous distribution of defects is treated by mimicked as a set of discrete states in some programs[11]. In AMPS and its derivatives such an energy distribution is treated by breaking the Gaussian into many bands of states as shown in **Fig. 1**. This allows the direct use of the results of the previous section. It follows from those results that the $n_{GD}(E_{GD})$, the number of electrons per volume occupying a given Gaussian distribution of defect states centered at the energy $E_{GD}$, is the sum of the electron populations of all the bands making up this Gaussian; i.e.,

$$n_{GD}(E_{GD}) = \sum_j \frac{N_j}{W} \int_{E_j-\frac{W}{2}}^{E_j+\frac{W}{2}} f_A(E)dE \qquad (20)$$

where *Nj* is the number of states per volume in band *j* and *Ej* is the center energy level of that band. It follows that the corresponding number of holes per volume in these states $p_{GD}(E_{GD})$ is given by

$$p_{GD}(E_{GD}) = \sum_j \frac{N_j}{W} \int_{E_j-\frac{W}{2}}^{E_j+\frac{W}{2}} f_D(E)dE \tag{21}$$

If the states in a given Gaussian are acceptor-like, they contribute $n_{GD}(E_{GD})$ to the quantity $n_t$ in Eq. (1). Alternatively, if the states in a given Gaussian are donor-like, they contribute $p_{GD}(E_{GD})$ to the quantity $p_t$ in Eq. (1). It follows from Eq. (8) that, regardless of whether the states are acceptor- or donor-like, the recombination traffic through a given Gaussian $R_{GD}(E_{GD})$ is given by

$$R_{GD}(E_{GD}) = \sum_j \frac{N_j}{W} \int_{E_j-\frac{W}{2}}^{E_j+\frac{W}{2}} \frac{(np-n_i^2)V_{th}\sigma_n\sigma_p dE}{\sigma_n(n+N_c e^{\frac{E-E_c}{\kappa T}})+\sigma_p(p+N_v e^{\frac{E_v-E}{\kappa T}})} \tag{22}$$

In this integration and summation, AMPS and its derivatives take the capture cross-sections to be constants for the whole Gaussian.  Again it is noted that it is the derivatives of these Gaussian defect state quantities with respect to $\psi$, $E_{fn}$ and $E_{fp}$ that are used in the Jacobian matrix element evaluations.

### 4.4.4 Band tails and Background Mid-gap states

In general there can be a distribution of localized defect states coming out of the conduction and valence bands. These distributions are a measure of the crystalline imperfection of the material structure and can be very significant for amorphous materials [1]. They reflect the character of the corresponding band states from which they came in the sense that those coming out of the valence band are donor-like states while those coming out of the conduction band are acceptor-like states. Often the distribution coming out of the conduction band fits the exponential decay dependence $N_{CT} = N_{CT0}e^{-E^C/E_a}$ where $E^C$ is measured positively down from the conduction band edge and $N_{CT0}$ as well as $E_a$ are fitting parameters. Correspondingly, the distribution coming out of the valence band often fits an exponential decay dependence $N_{VT} = N_{VT0}e^{-E^V/E_d}$ where $E^V$ is measured positively up from the valence band edge and $N_{VT0}$ as well as $E_d$ are fitting parameters. Such exponential distributions are termed Urbach tails. AMPS and its derivatives assume any band tails present fit the Urbach model and divide these exponential distributions into numerous bands of states as seen in **Fig. 2**. Using this series of bands to represent the Urbach tail, the number of electrons occupying the conduction band tail per volume $n_{CTD}$ is computed from

$$n_{CTD} = \sum_i \frac{N_{CTi}}{W} \int_{E_i-\frac{W}{2}}^{E_i+\frac{W}{2}} f_A(E)dE \tag{23}$$

Similarly the number of holes occupying the valence band tail per volume $p_{VTD}$ is given by

$$p_{VTD} = \sum_i \frac{N_{VTi}}{W} \int_{E_i-\frac{W}{2}}^{E_i+\frac{W}{2}} f_D(E)dE \tag{24}$$

The quantities $N_{VTi}$ and $N_{CTi}$ are the average state densities for the $i^{th}$ donor-like and acceptor-like banded state, respectively. They satisfy

$$N_{VTi} = \int_{E_i-\frac{W}{2}}^{E_i+\frac{W}{2}} N_{VT0} e^{-(E-E_V)/E_d} \, dE \tag{25}$$

and

$$N_{CTi} = \int_{E_i-\frac{W}{2}}^{E_i+\frac{W}{2}} N_{CT0} e^{-(E_c-E)/E_a} \, dE \tag{26}$$

which results in

$$N_{VTi} = N_{VT0} E_d (e^{-(E_i-E_V-W/2)/E_d} - e^{-(E_i-E_V+W/2)/E_d}) \tag{27}$$

$$N_{CTi} = N_{CT0} E_a (e^{(E_i-E_c+W/2)/E_a} - e^{(E_i-E_c-W/2)/E_a}) \tag{28}$$

Since the conduction band tail states are acceptor-like, they contribute $n_{CTD}$ to the quantity $n_t$ in Eq. (1). Since the valence band tail states are donor-like, they contribute $p_{VTD}$ to the quantity $p_t$ in Eq. (1).

The recombination traffic through the conduction band tail $R_{CTD}$ is given by

$$R_{CTD} = \sum_j \frac{N_{CTj}}{W} \int_{E_j-\frac{W}{2}}^{E_j+\frac{W}{2}} \frac{(np-n_i^2) V_{th} \sigma_n \sigma_p \, dE}{\sigma_n(n+N_c e^{\frac{E-E_c}{\kappa T}}) + \sigma_p(p+N_v e^{\frac{E_v-E}{\kappa T}})} \tag{29}$$

Correspondingly, the recombination traffic through the valence band tail $R_{VTD}$ is given by

$$R_{VTD} = \sum_j \frac{N_{VTj}}{W} \int_{E_j-\frac{W}{2}}^{E_j+\frac{W}{2}} \frac{(np-n_i^2) V_{th} \sigma_n \sigma_p \, dE}{\sigma_n(n+N_c e^{\frac{E-E_c}{\kappa T}}) + \sigma_p(p+N_v e^{\frac{E_v-E}{\kappa T}})} \tag{30}$$

AMPS and its derivatives assume the tail capture cross-sections do not vary with energy.

The band tails may decay as they penetrate into the energy gap to a point where they become dominated by background mid-gap defect states as seen in **Fig. 2**. In **Fig. 2** these background states are assumed to be of constant density and in that case would be modeled using the tools developed in the banded defects section.

We note that it is the derivatives of these tail and background mid-gap defect state quantities with respect to $\psi$, $E_{fn}$ and $E_{fp}$ that are used in the Jacobian matrix element evaluations employed below.

### 4.4.5 Amphoteric states

Some defect states have an amphoteric nature; i.e., they are multi-particle, multivalent, and have various charged states. We consider the situation in which this type of state is positive when unoccupied by two electrons, neutral when occupied by an electron, and negative when occupied by two electrons [1, 7]. An excellent example of such defect states is found in the dangling bonds of amorphous silicon. For a localized amphoteric state of this type, there are two energy levels located at *E* and *E+U*. The correlation energy *U* is caused by the repulsive coulomb interaction and the nearest neighbor distortion, and is generally accepted to be positive for the dangling bond states of a-Si:H [12]. The level *E* is populated in the transition between the positive and neutral charged states, and the level *E+U* is populated in the transition between the neutral and negative charged states. These transitions are performed through capture and emission processes of electrons and holes at each energy level (shown in **Fig. 3**). Two models that evaluate the recombination and trapping statistics for amphoteric states by different treatments are discussed below.

### 4.4.5.1 Applying SRH statistics to Amphoteric States

As seen in **Fig. 3**, the recombination traffic at each energy level of the amphoteric state is similar to the one of the SRH recombination process. Instead of characterizing a single amphoteric state by two coupled transition levels, a simplistic way to try to mimic the amphoteric state is by creating a pair of decoupled levels in the band gap which we will call SRH states: a donor-like state with a Gaussian distribution (and corresponding capture-cross-sections) at level $E_{ad}$ = *E* and an acceptor-like state with a Gaussian distribution (and corresponding capture-cross-sections) at level $E_{aa}$ = *E+U*. If degeneracy factors of multivalent states are considered for this amphoteric state, the energy levels of the two uncorrelated SRH states should be shifted slightly from the correlated transition levels of the amphoteric state for a better approximation. This results in [13]:

$$E_{ad} = E - kT \ln 2 \tag{31a}$$

$$E_{aa} = E + U + kT \ln 2 \tag{31b}$$

The adjustment of energy levels allows the net charge of the SRH states to be identical to that of the amphoteric state at thermodynamic equilibrium. The electrons populating and depopulating at these SRH levels are designed to represent the charge transitions occurring at the amphoteric state. The equations described in Section 3 can be used here to analyze the recombination statistics. A reasonable assumption is to take state densities to be the same density and to equal that of the entire amphoteric state [14].

This simplified approach for modeling amphoteric states is used in AMPS and has been applied to simulate the effects of dangling bonds in a-Si solar cells [15, 16]. However, this decoupled recombination model neglects the coupled nature within the transition levels of amphoteric state, and could result in some degree of error. The inaccuracy introduced by this method has been discussed in several articles [14, 17, 18]. However, it is commonly agreed that the simplified method is an excellent approximation when the correlation energy is positive and satisfies $U \gg kT$, and the capture cross-sections of the neutral state are much smaller than the ones of charged states. And the separation energy of $E_{ad}$ and $E_{aa}$ should be kept always as $U+2kTln2$ when amphoteric states are approximated by this approach.

### 4.4.5.2 Applying Sah-Shockley statistics to Amphoteric States

A more precise model for characterizing the amphoteric state is based on the recombination and trapping statistics developed by Sah and Shockley [19]. In this model, the correlated nature within transition levels is considered, and the amphoteric state is treated as a whole.

For the amphoteric state with a density of $N_{AmD}$, the recombination rates of electrons and holes at the transition level $E$ are denoted by $R_n^E$ and $R_p^E$ [20]:

$$R_n^E = n\sigma_n^E V_{th} N_{AmD} F^+ - e_n^E N_{AmD} F^0 \tag{32}$$

$$R_p^E = p\sigma_p^E V_{th} N_{AmD} F^0 - e_p^E N_{AmD} F^+ \tag{33}$$

where $\sigma_n^E$ and $\sigma_p^E$ are cross-section capture areas of electrons and holes for the $E$ level, $e_n^E$ and $e_p^E$ are emission coefficients of electrons and holes for the $E$ level. $F^+$, $F^0$ are occupation probabilities of positive charged state, neutral charged state, respectively. Eqs. (32) and (33) reflect the transition process between the zero-electron state and the one-electron state. Similar equations can be applied to electron recombination rate $R_n^{E+U}$ and hole recombination rate $R_p^{E+U}$ at the transition level $E+U$. The quantities $e_n^E$ and $e_p^E$ can be obtained by applying the detailed balance rule[21], According to $R_n^E = R_p^E$, $R_n^{E+U} = R_p^{E+U}$, and $F^+ + F^0 + F^- = 1$ at steady condition, where $F^-$ is the occupation probability of negative charged state, the occupation probabilities are obtained as[17]:

$$F^+(E) = \frac{P^E P^{E+U}}{N^E P^{E+U} + P^E P^{E+U} + N^E N^{E+U}}$$

$$F^0(E) = \frac{N^E P^{E+U}}{N^E P^{E+U} + P^E P^{E+U} + N^E N^{E+U}}$$

$$F^-(E) = \frac{N^E N^{E+U}}{N^E P^{E+U} + P^E P^{E+U} + N^E N^{E+U}} \tag{34}$$

where $P^E = pV_{th}\sigma_p^E + e_n^E$, $P^{E+U} = pV_{th}\sigma_p^{E+U} + e_n^{E+U}$, $N^E = nV_{th}\sigma_n^E + e_p^E$, $N^{E+U} = nV_{th}\sigma_n^{E+U} + e_p^{E+U}$. Hence, the amphoteric state with a density of $N_{AmD}(E)$ per volume per energy contributes $N_{AmD}(E)F^-(E)$ to $n_t$ in Eq. (1), and $N_{AmD}(E)F^+(E)$ to $p_t$ in Eq. (1).

The total recombination rate through the amphoteric state, $R_{AmD}(E)$ is the sum of the recombination traffic of two energy levels, $R_n^E + R_n^{E+U}$, or $R_p^E + R_p^{E+U}$. And $R_{AmD}(E)$ is found to be[17]:

$$R_{AmD}(E) = \frac{N_{AmD}(E)V_{th}^2(np - n_i^2)(\sigma_n^E \sigma_p^E P^{E+U} + \sigma_n^{E+U} \sigma_p^{E+U} N^E)}{N^E P^{E+U} + P^E P^{E+U} + N^E N^{E+U}} \tag{35}$$

Multiple uses of Eqs. (34) and (35) with different $N_{AmD}$, $E$, and capture cross-section values allow very general amphoteric defect state distributions to be constructed in the band gap. Again we stress that it is the derivatives of these amphoteric defect state quantities with respect to $\psi$, $E_{fn}$ and $E_{fp}$ that are used in the Jacobian matrix element evaluations employed below.

As seen in Eqs. (34) and (35), the calculation of occupied charges and recombination traffic very different from the SRH method described in the previous section. However, in some specific conditions (discussion in section 5.1), both approaches produce close results. Currently AMPS and its derivatives use the SRH method. We intend to have the precise Sah-Shockley model for the amphoteric states incorporated in the later versions of AMPS derivatives.

## 5. AN EXAMPLE OF SOLVING THE SYSTEM NUMERICALLY: THE AMPS APPROACH

The mathematical system defining solar cell operation at steady state includes Eqs (1) to (3) and also the equations of section IV, as appropriate. In solving this mathematical system numerically, the device being modeled is discretized into *N* regions giving rise to *N-1* internal nodes and two contact nodes thereby giving a total of *N+1* nodes. The code defines the state of the device by determining $\psi$, $E_{fn}$ and $E_{fp}$ at each node. Solving for these at some $i^{th}$ node in the domain is accomplished by writing Poisson's and the two continuity equations as differences $F_1$, $F_2$ and $F_3$ [22]:

$$F_1(\psi, E_{fn}, E_{fp}) = \frac{d}{dx}(\varepsilon \frac{d}{dx}\psi) - q(p - n + N_d^+ - N_a^- + p_t - n_t) = 0 \tag{36}$$

$$F_2(\psi, E_{fn}, E_{fp}) = \frac{dJ_n}{dx} - q(R - G) = 0 \tag{37}$$

$$F_3(\psi, E_{fn}, E_{fp}) = \frac{dJ_p}{dx} - q(G - R) = 0 \tag{38}$$

The three difference functions at the *(N+1)* nodes must be simultaneously driven to zero to obtain the exact solutions for $\psi$, $E_{fn}$, and $E_{fp}$ at these points.

In the numerical solution approach used in AMPS, the difference functions for the boundaries and the $F_1$, $F_2$ and $F_3$ differences for the *N-1* internal nodes are thought of as functions of independent variables $\psi$, $E_{fn}$ and $E_{fp}$ at each node. Actually these differences will be functions of $\psi$, $E_{fn}$ and $E_{fp}$ values at nearby nodes too. This will occur due to the spatial finite differences that are utilized in AMPS to calculate the derivatives involved in Equations (36)-(38). For example, for the $i^{th}$ internal node:

$$F_1(i) = A_1\psi_{i-1} - (A_1 + A_2)\psi_i + A_2\psi_{i+1} - q(p - n + N_d^+ - N_a^- + p_t - n_t) \tag{39}$$

$$F_2(i) = \frac{2(J_n(i) - J_n(i-1))}{(x_{i+1} - x_{i-1})} - q(R - G) \tag{40}$$

$$F_3(i) = \frac{2(J_p(i) - J_p(i-1))}{(x_{i+1} - x_{i-1})} - q(G - R) \tag{41}$$

where [10]:

$$A_1 = \frac{4\varepsilon_{i-1}\varepsilon_i}{(x_i - x_{i-1})(x_{i+1} - x_{i-1})(\varepsilon_{i-1} + \varepsilon_i)} \quad \text{and} \quad A_2 = \frac{4\varepsilon_i\varepsilon_{i+1}}{(x_{i+1} - x_i)(x_{i+1} - x_{i-1})(\varepsilon_i + \varepsilon_{i+1})} \tag{42}$$

$$J_n(i) = \frac{q\mu_n N_c(\psi_{i+1} - \chi_{i+1} - \psi_i + \chi_i)}{(x_{i+1} - x_i)(e^{\frac{\psi_{i+1} - \chi_{i+1}}{\kappa T}} - e^{\frac{\psi_i - \chi_i}{\kappa T}})} (e^{\frac{E_{fn_{i+1}}}{\kappa T}} - e^{\frac{E_{fn_i}}{\kappa T}}) \tag{43}$$

$$J_p(i) = \frac{q\mu_p N_v(\psi_{i+1} - \chi_{i+1} - Eg_{i+1} - \psi_i + \chi_i + Eg_i)}{(x_{i+1} - x_i)(e^{\frac{-\psi_{i+1} + \chi_{i+1} + Eg_{i+1}}{\kappa T}} - e^{\frac{-\psi_i + \chi_i + Eg_i}{\kappa T}})} (e^{\frac{-E_{fp_{i+1}}}{\kappa T}} - e^{\frac{-E_{fp_i}}{\kappa T}}) \tag{44}$$

The Scharfetter-Gummel discretization method [23] has been applied here to the generalized drift-diffusion expressions for $J_n(i)$ and $J_p(i)$.

Obtaining the exact solutions for $\psi$, $E_{fn}$, and $E_{fp}$ at the nodes requires that the difference functions at the *(N+1)* nodes be simultaneously driven to zero. In AMPS and its derivatives the Newton-Raphson solution technique [2, 24] is employed for this effort with $\psi$, $E_{fn}$ and $E_{fp}$ as the independent variables.

In solving these equations for $\psi$, $E_{fn}$ and $E_{fp}$ at each node a key task becomes evaluating the Jacobian matrix elements arising from the use of the Newton-Raphson method [24]. In principle, developing this Jacobian matrix requires partial derivatives of every difference function with respect to every independent variable, which means that the size of the matrix in our system is *3(N+1)* by *3(N+1)* and the elements are composed of $\frac{\partial Difference_{i,j}}{\partial var_{k,l}}$, where *i* varies over the nodes from 1 to *N+1* and *j* varies over the three difference statements at each node, and *k* varies over the nodes from 1 to *N+1* and *l* varies over the three difference statements at each node and *l* denotes $\psi$, $E_{fn}$ and $E_{fp}$ at each node. Fortunately, Eq. (39)–(44) show the differences for the $i^{th}$ internal node that are evaluated in AMPS codes by using the variables at the $i^{th}$ and neighboring *(i-1)*$^{th}$ and *(i+1)*$^{th}$ nodes. Similarly, the difference statements at the boundaries only involve the boundary and the immediately adjacent nodes. As a consequence, the Jacobian matrix is simplified to a banded matrix with a bandwidth of three, and in each iteration the variation of $\psi$, $E_{fn}$ and $E_{fp}$ at each node are reasonably easily to be solved by using the Lower Upper decomposition method [24].

## 6. CONCLUSIONS

In this paper we argue that the impact of band-edge-property variations with position and gap state effects can become prominent in solar cell devices and that, with the computer power available today, should be included in any solar cell numerical performance modeling. To assist in this endeavor, we review the methodology for including in transport modeling the effective forces arising from band-edge-property variations with position. Further we catalogue and review the gap state effects possible in solar cell structures and their influence on transport. Gap state population and recombination models, which quantify trapped charge and recombination rates arising from the dopant states, discrete, banded and Gaussian gap localized states, Urbach band tail, and background mid-gap states, are established. This whole set of effective forces and band gap state impacts equations is used in a difference version of the Eq. (1)-(3) set which is then discretized as shown in the example of Eq. (39)~(41). Obtaining a numerical solution is exemplified by reviewing the approach used in AMPS, which employs the Newton-Raphson method. This comprehensive model has been implemented in the AMPS family of codes, and has been utilized to analyze the effects of various defects to solar cell characteristics.

## ACKNOWLEDGEMENT

The authors greatly appreciate the supports from Prof. Yun Sun at Nankai University, China, and Prof. Angus Rockett at Colorado School of Mines.

## APPENDIX

To evaluate Eq. (18), the following expression may be written:

$$p_{BD}(E_{BD}) = N_{BD} \int_{E_{BD}-\frac{W}{2}}^{E_{BD}+\frac{W}{2}} \frac{(\sigma_p p + c_1 e^{\frac{x-E_g}{\kappa T}}) dx}{c_1 e^{\frac{x-E_g}{\kappa T}} + c_2 e^{\frac{-x}{\kappa T}} + \sigma} \quad \text{(A1)}$$

where $c_1 = \sigma_n N_c$, $c_2 = \sigma_p N_v$, $\sigma = \sigma_n n + \sigma_p p$, $x = E - E_v$. The integral in Eq. (A1) is expanded as,

$$\frac{1}{2} \int_{E_{BD}-\frac{W}{2}}^{E_{BD}+\frac{W}{2}} \frac{(2c_1 e^{\frac{x-E_g}{\kappa T}} + \sigma) dx}{c_1 e^{\frac{x-E_g}{\kappa T}} + c_2 e^{\frac{-x}{\kappa T}} + \sigma} + \frac{\sigma_p p - \sigma_n n}{2} \int_{E_{BD}-\frac{W}{2}}^{E_{BD}+\frac{W}{2}} \frac{dx}{c_1 e^{\frac{x-E_g}{\kappa T}} + c_2 e^{\frac{-x}{\kappa T}} + \sigma} \quad \text{(A2)}$$

Denote the first integral term as $P_{term}$, and the second integral as $R_{term}$. $P_{term}$ is given by,

$$P_{term} = \begin{cases} \kappa T \ln(c_1 e^{\frac{x-E_g}{\kappa T}} + c_2 e^{\frac{-x}{\kappa T}} + \sigma) + x \Big|_{E_{BD}-\frac{W}{2}}^{E_{BD}+\frac{W}{2}} & \text{if } \Delta \neq 0 \\ 2\kappa T \ln(2c_1 e^{\frac{x-E_g}{\kappa T}} + \sigma) \Big|_{E_{BD}-\frac{W}{2}}^{E_{BD}+\frac{W}{2}} & \text{if } \Delta = 0 \end{cases} \quad (A3)$$

where $\Delta = 4c_1 c_2 e^{\frac{-E_g}{\kappa T}} - \sigma^2$. $R_{term}$ is given by,

$$R_{term} = \begin{cases} \dfrac{2\kappa T}{\sqrt{\Delta}} \arctan \dfrac{2c_1 e^{\frac{x-E_g}{\kappa T}} + \sigma}{\sqrt{\Delta}} \Big|_{E_{BD}-\frac{W}{2}}^{E_{BD}+\frac{W}{2}} & \text{if } \Delta > 0 \\ \dfrac{-2\kappa T}{2c_1 e^{\frac{x-E_g}{\kappa T}} + \sigma} \Big|_{E_{BD}-\frac{W}{2}}^{E_{BD}+\frac{W}{2}} & \text{if } \Delta = 0 \\ \kappa T \ln\left(\dfrac{2c_1 e^{\frac{x-E_g}{\kappa T}} + \sigma - \sqrt{-\Delta}}{2c_1 e^{\frac{x-E_g}{\kappa T}} + \sigma + \sqrt{-\Delta}}\right) \Big|_{E_{BD}-\frac{W}{2}}^{E_{BD}+\frac{W}{2}} & \text{if } \Delta < 0 \end{cases} \quad (A4)$$

Based on the evaluation of $P_{term}$ and $R_{term}$, Eqs. (17)-(19) can be expressed as:

$$n_{BD}(E_{BD}) = N_{BD}W - p_{BD}(E_{BD}) \tag{A5}$$

$$p_{BD}(E_{BD}) = N_{BD}\left(\frac{1}{2}P_{term} + \frac{\sigma_p p - \sigma_n n}{2}R_{term}\right) \tag{A6}$$

$$R_{BD}(E_{BD}) = N_{BD}(np - n_i^2)V_{th}\sigma_n\sigma_p R_{term} \tag{A7}$$

And their derivatives required by the Jacobian matrix can be deduced from Eqs. (A5)-(A7).

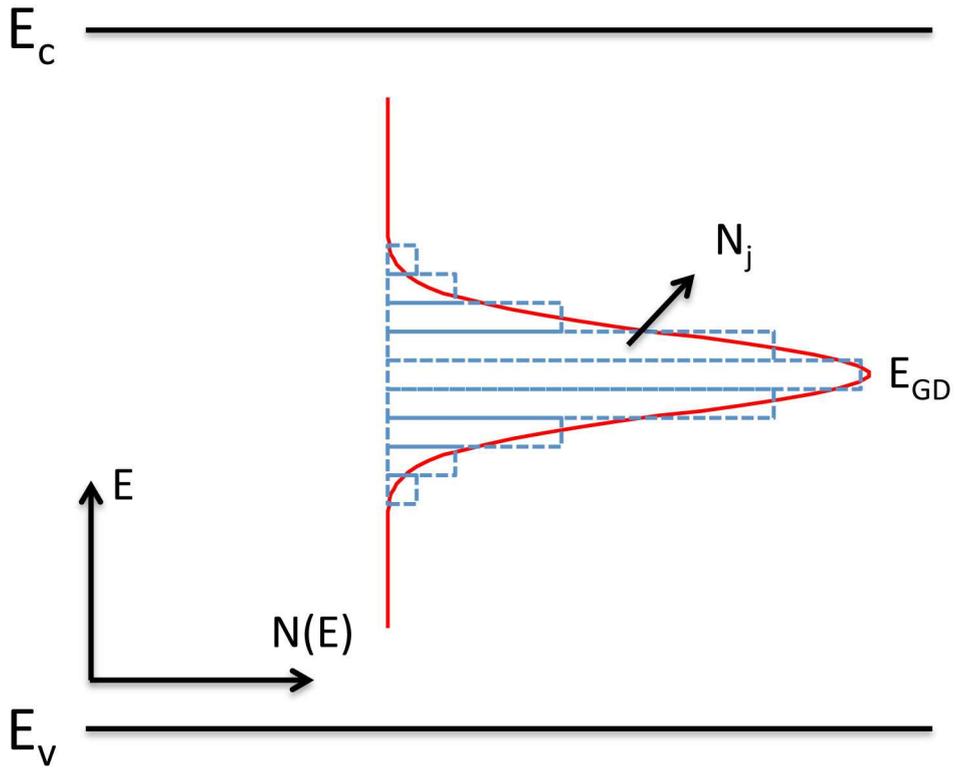

**Figure 1. A Gaussian density of states versus energy distribution centered at energy $E_{GD}$. $N(E)$ stands for the density of states per energy. The band $N_j$ is an example of an energy band of width $W$ centered at the energy $E_j$ in this distribution.**

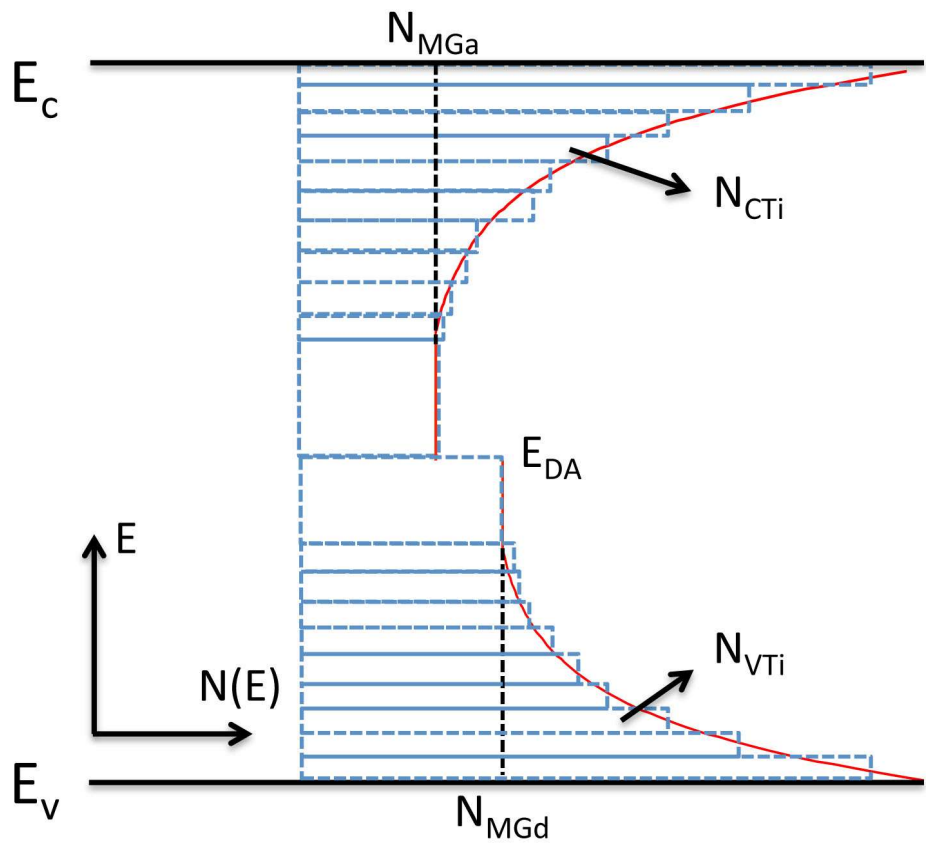

**Figure 2. Band tail defect states and background Mid-gap states $N_{MGa}$ and $N_{MGd}$.** The energy $E_{DA}$ is the switch-over energy level for acceptor-like and donor-like states, $N_{MGa}$ and $N_{MGd}$ are densities of mid-gap acceptor-like and donor-like states, respectively.

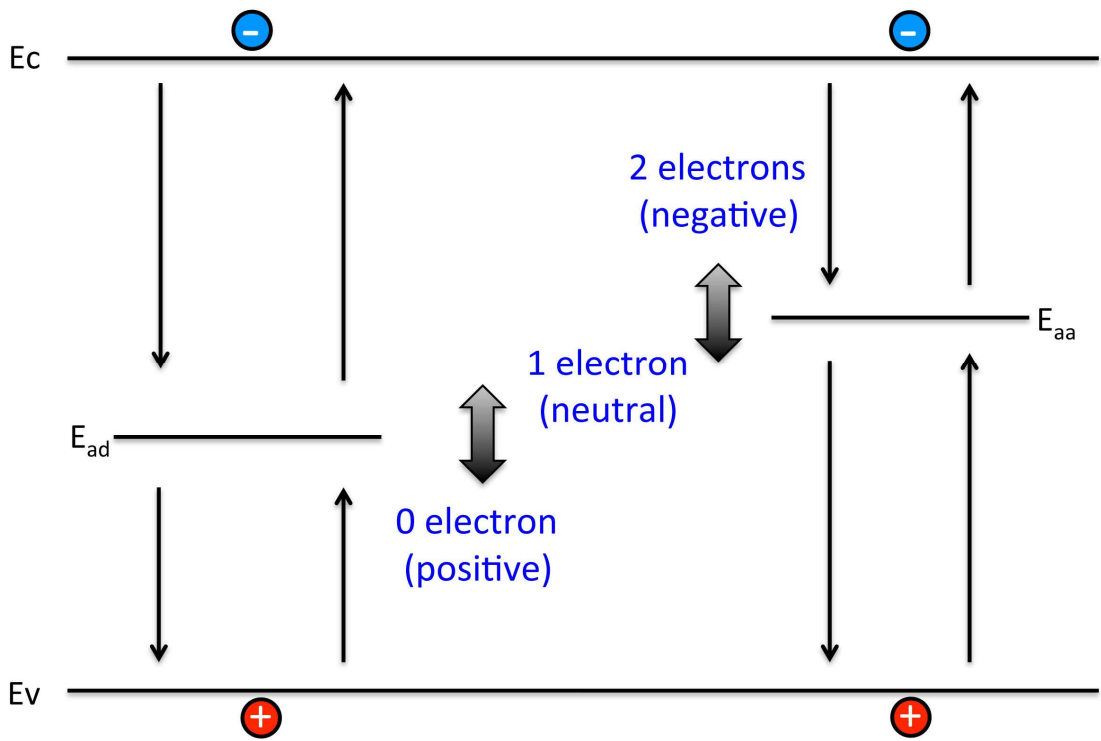

**Figure.3 Illustration of charged state transitions on an amphoteric state.**